\documentclass[aip, jmp, amsmath,amssymb,]{revtex4}

\usepackage{graphicx}
\usepackage{dcolumn}
\usepackage{bm}

\usepackage[utf8]{inputenc}
\usepackage[T1]{fontenc}
\usepackage{mathptmx}
\usepackage{etoolbox}
\usepackage{enumerate}
\usepackage{amsthm}
\usepackage{color}
\usepackage{graphicx}
\usepackage{epstopdf}
\usepackage{float}
\usepackage{thmtools}
\usepackage{thm-restate}
\usepackage{hyperref}
\usepackage{cleveref}
\usepackage{tabularx}
\usepackage{tikz,amstext}

\newtheorem{thm}{Theorem}[section]
\newtheorem{cor}[thm]{Corollary}
\newtheorem{lem}[thm]{Lemma}
\newtheorem{prop}[thm]{Proposition}

\newtheorem{notation}{Notation}[section]

\newtheorem{defn}{Definition}[section]
\newtheorem{rem}{Remark}[section]

\newcommand{\un}{1\mkern -4mu{\rm l}}
\newcommand{\A}{\mathcal{A}}

\newcommand{\M}{\mathcal{M}}

\newcommand{\R}{\mathbb{R}}
\newcommand{\N}{\mathbb{N}}

\DeclareMathOperator{\tr}{tr}

\makeatletter
\def\@email#1#2{%
 \endgroup
 \patchcmd{\titleblock@produce}
  {\frontmatter@RRAPformat}
  {\frontmatter@RRAPformat{\produce@RRAP{*#1\href{mailto:#2}{#2}}}\frontmatter@RRAPformat}
  {}{}
}%
\makeatother
\begin{document}

\preprint{AIP/123-QED}

\title[Limit distribution of partial transposition of block random matrices]{Limit distribution of partial transposition of block random matrices}
\author{Zhi Yin}
\altaffiliation{School of Mathematics and Statistics, Central South University, China}%
\author{Liang Zhao}
\altaffiliation{School of Mathematics, Harbin Institute of Technology, China}

\date{\today}

\begin{abstract}
It is well known that, under some assumptions, the limit distribution of random block matrices and their partial transposition converges to 
the distributions of random variables in some noncommutative probability space. 
Using free probability theory, we obtain the relation between the free cumulants of the corresponding random variables. 
As an application, we are able to derive a new family of co-completely positive and $k$-positive maps by using the Wishart ensemble.  
\end{abstract}

\maketitle


\section{Introduction}

The random matrix theory \cite{AGZ}, which dates back to the work of Wigner and Wishart, has a long history in physics. Yet, it becomes a powerful tool in quantum information in recent years. Many progress has been made by using this tool \cite{BC, CHN2016, CYZ, IN07, AS14, FM13, INA14, INA19, ILS20}. A suitable random matrix model may provide an extra room, thus allowing plenty of mathematical tools to be involved, which are beneficial for addressing the problems.

The following (random) block matrix model is important for the quantum entanglement theory \cite{HHHH09}. Let $X_{nN} \in \M_n(\mathbb{C}) \otimes \M_N(\mathbb{C})$ be a Hermitian (random) matrix, and define 
$$X_{nN}^\Gamma : = \un_n \otimes \Gamma \left( X_{nN} \right)$$
be the partial transposition of $X_{nN},$ where $\Gamma: \M_N(\mathbb{C}) \rightarrow \M_N(\mathbb{C})$ is the transpose map. The motivation for considering this model comes from 
the famous positive partial transpose (PPT) criteria: any PPT state is non-distillable, 
thus any entangled PPT state is bound entangled \cite{HHH98}. 
Hence, if one considers $X_{nN}$ as a (non-normalized) quantum state, determining the eigenvalue distribution of
$X_{nN}^\Gamma$ can aid in determining the PPT property of $X_{nN}.$
In the large dimension, the limit distribution of above mentioned random matrix models can be described using random variables in the framework of free probability \cite{D94, NS2006, MS17}. More precisely, suppose that $X_{nN}$ is Haar unitary invariant (e.g. $X_{nN}$ is GUE or Wishart ensemble), there is a noncommutative probability space $(\mathcal{A}, \varphi)$ and random variable $x$ (resp. $\tilde{x}^\Gamma$) in $(\mathcal{A}, \varphi)$ with distribution $\mu$ (resp. $\mu^\Gamma$), such that the eigenvalue distribution of $X_{nN}$ (resp. $X_{nN}^\Gamma$) converges to $\mu$ (resp. $\mu^\Gamma$). We refer to Section \ref{sec:II} for more details in free probability theory. A natural question is how to determine $\mu^\Gamma$ via $\mu$ and we refer to \cite{G12, BN13} for more details. 
In \cite{ANV16, IN2018}, the authors explicitly computed the R-transform of $\mu^\Gamma$ in terms of the R-transform of $\mu$ by using the theory of operator-valued free probability\cite{RS1998, DJ18, RT17, RDN02, CJ15, ND92}. In this paper, we revisit this question, and we obtain a simple formula for the free cumulants of $\tilde{x}^\Gamma$ (see Theorem \ref{thm:1}) by only using some simple combinatoric techniques. 

Moreover, as an application, we explicitly compute the following model: 
$$X_{nN} = \un_{nN} + \alpha W_{nN},$$
where $\alpha$ is a parameter and $X_{nN}$ is an $nN \times nN$ random Wishart matrix. The motivation for considering this model is to tackle the following NPT problem:
{\it Find a bound entangled state with non-positive partial transposition (NPT).} Many progress have been done after the proposal of the problem \cite{DSSTT2000, DCLB2000, BR03, PPHH2010}, however, it is still open. We note that the NPT problem can be solved by studying the 2-positivity of the tensor product of co-completely maps \cite{DSSTT2000} (see \cite{CTY18} for recent progress). Namely, the existence of NPT bound entanglement is equivalent to the following: { \it Find a linear map $\Phi$ such that
(i) $\Phi$ is co-completely positive; (ii) $\Phi^{\otimes r}$ is 2-positive for all $r\geq 1$.}
More generally,  one can study the tensor-stable positivity of linear maps \cite{MRW15, H06}, and it was shown that the non-trivial tensor-stable positivity implies the NPT bound entanglement \cite{MRW15}. 

In this work, let $\Phi_N: \M_n(\mathbb{C}) \rightarrow \M_N(\mathbb{C})$ be a linear map such that $\Phi_N$ is the Choi map of $X_{nN}$. By choosing proper parameters one has:
almost surely as $N \to \infty$,
\begin{enumerate}[{\rm(a)}]
\item $\Phi_N$ is co-completely positive, while $\Phi_N$ is not completely positive;
\item for any given integer $k \geq 1$, $\Phi_N$ is $k$-positive.
\end{enumerate}
Therefore, a direct corollary is that we find a new family of 1-copy non-distillable states with non-positive partial transposition (see Propositions \ref{prop:PPT} and \ref{prop:k-positive}). However, in order to solve the NPT problem, one has to 
consider the $k$-positivity of the tensor product of $\Phi_N.$

The rest of the paper is structured as follows: Section \uppercase\expandafter{\romannumeral2} introduces the necessary notions of free probability theory. Strongly convergence of random
matrices are recalled in Section \uppercase\expandafter{\romannumeral3}. In Section \uppercase\expandafter{\romannumeral4}, we revisit the strongly convergence of our random matrix model
and its partial transposition, and give an alternative proof based on the resuals of Haagerup-Thorbj{\o}rnsen and Collins-Male. Section \uppercase\expandafter{\romannumeral5} provides the
limit distribution of the partially transposed random matrix under a combinatoric method and a detailed example. In Section \uppercase\expandafter{\romannumeral6}, we study the parameter
range when the matrix model satisfied the co-complete positivity and $k-$positivity conditions.


\section{Brief introduction to free probability}\label{sec:II}

We will try to briefly introduce the basic idea of free probability. And we refer to  \cite{NS2006, MS17} for more details. Free probability, which is a quantum analogy of classical probability, is an interdispline of functional analysis, operator algebra, combinatorics, et.al. It was introduced by Voiculescu \cite{D94} to tackle the famous 
"isomorphism of free group factors" problem in operator algebra. In free probability, the random variables are some operators that sit in a $*$-algebra $\mathcal{A}$ with a normal faithful tracial state $\varphi$.  And, instead of "classical independence", "free independence" takes a central role.

\vspace{2mm}

\noindent {\it Noncommutative probability spaces and freely independence}--A noncommutative probability space $(\mathcal{A},\varphi)$ is an unital algebra $\mathcal{A}$ over $\mathbb{C}$, with an unital linear functional
\[\varphi:\mathcal{A}\to\mathbb{C}; \;  \varphi(\mathbf{1}_{\mathcal{A}})=1.\]
The element $x\in\mathcal{A}$ is called noncommutitive random variable, and it is called centred if $\varphi(x)=0$. 

Let $\mathcal{A}_{1},\mathcal{A}_{2},\ldots,\mathcal{A}_{r}$ be subsets of $\mathcal{A}$. Denote  $alg(\mathcal{A}_{i})$ by the algebra generated by $\mathcal{A}_i$. If for any centred elements $x_{j}\in alg(\mathcal{A}_{i_{j}})$, $j=1,\ldots, k$, 
\begin{equation}\label{eq:free-independent}
\varphi(x_{1}\cdots x_{k})=0
\end{equation}
whenever we have  $i_{j}\ne i_{j+1}$ for $j=1,\ldots, r-1$, then we say $\mathcal{A}_{1},\mathcal{A}_{2},\ldots,\mathcal{A}_{r}$ is freely independent.

Suppose that $\mathcal{A}$ is a $\ast$-algebra and $\varphi$ is positive, i.e., $\varphi(xx^{\ast})\geq0$ for all $x \in \mathcal{A}.$
Then $(\mathcal{A},\varphi)$ is called a $\ast$-probability space. Additionally, $\varphi$ is tracial if $\varphi(xy)=\varphi(yx)$ for all $x, y\in\mathcal{A}$. And $\varphi$ is faithful if have the implication
\[x\in\mathcal{A},\ \varphi(x^*x)=0 \; \Rightarrow \; x=0.\]
Moreover, suppose that $\mathcal{A}$ is a $C^*$-algebra equipped with a norm $\|\cdot\|,$ $(\mathcal{A}, \varphi)$ is called a $C^*$-probability space.
Here are two examples of $*$-probability spaces: 
\begin{enumerate}[{\rm (i)}]
\item Let $\mu$ be a probability space supporting on $\R,$ then the function space $L^\infty(\R, \mu),$ together with the expectation $\mathbb{E}$ which is defined by $\mathbb{E} [f] := \int_{\R} f(t) d \mu(t)$, is a $*$-probability space.
\item Let $\M_N(\mathbb{C})$ be the algebra of $N \times N$ matrices, and ${\rm tr}_N = {\rm Tr}/N$, then $(\M_N(\mathbb{C}), {\rm tr}_N)$ is a $*$-probability space. 
\end{enumerate}

 \vspace{2mm}
 
 \noindent {\it Free cumulants and free convolution}--Let $(\A,\varphi)$ be a $*$-probability space, and $x \in \A.$ $\varphi(x^{\epsilon_{1}}\cdots x^{\epsilon_{k}})$ is called the $*$-moments of $x$, where $k\geq0$ and $\epsilon_{1},\cdots,\epsilon_{k}\in\{1, *\}.$ The $*$-distribution of $x$ is the linear functional $P \rightarrow \varphi \left[ P(x, x^*)\right]$ on the set of polynomials in 2 noncommutative indeterminates. 
  
In general case, for a family of random variables $x_{1},\ldots, x_{p}$, let us define 
\[\varphi_{p}(x_{1},\ldots, x_{p}):=\varphi(x_{1}\cdots x_{p}). \]
Denote $NC(p)$ by the set of non-crossing partitions for the index set $[1, \ldots, p].$ For any partition $\pi\in NC(p)$, define
\[\varphi_{\pi}[x_{1},\ldots,x_{p}]:=\prod_{V\in\pi}\varphi(V)[x_{1},\cdots,x_{p}],\]
where
\[\varphi(V)[x_{1},\cdots,x_{p}]:=\varphi_{|V|}(x_{i_{1}},\cdots,x_{i_{r}})\ \ \text{for}\ V =(i_{1},\cdots, i_{r}).\]
Hence $(\varphi_{\pi})_{\pi \in NC(p)}$ are multilinear functionals on $\A.$

The free cumulants $(\kappa_{\pi})_{\pi\in NC(p)}$ are the unique multilinear functionals on $\mathcal{A}$ such that 
\begin{equation}\label{eq:moment-cumulant}
\varphi(x_{1}\cdots x_{p})=\sum_{\pi\in NC(p)}\kappa_{\pi}[x_{1},\cdots, x_{p}].
\end{equation}
For any integer $p\geq 0,$ we will write 
$$\kappa_p[x_1, \ldots, x_p]:= \kappa_{\mathbf{1}_p}[x_1, \ldots, x_p],$$
where $\mathbf{1}_{p}$ is the identity permutation in the index set $[1, \ldots, p].$

In particular, suppose $x$ is a self-adjoint element in $(\mathcal{A},\varphi)$. If there is a probability measure $\mu$ which is compactly supported on $\R,$ such that for any integer $p\geq 0,$ we have
\[\varphi(x^{p})=\int_{\mathbb{R}}t^{p}\mathrm{d}\mu(t).\] 
Then $\mu$ is called the distribution of $x$. The Cauchy-transform $G_\mu(z)$ and R-transform $R_\mu(z)$ of $\mu$ (respect to $x$) are given by
\begin{equation}
G_{\mu}(z)=\sum_{p=0}^{\infty}\frac{\varphi(x^{p})}{z^{p+1}},\  R_{\mu}(z)=\sum_{p=0}^{\infty}\kappa_{p+1}[x, \ldots, x] z^{p}.
\end{equation}

Given two (self-adjoint) random variables $x$ and $y,$ it is natural to consider the probability distribution of their various combinations, e.g., the sum and the product. In the classical theory, it is well-known that the probability distribution of $x+y,$ where additionally we require $x$ and $y$ are independent, is the convolution of $\mu_x$ and $\mu_y.$ This result surely has the following counterpart in free probability: if $x$ and $y$ are freely independent, then the probability distribution of $x+y$ is the free convolution of $\mu_x$ and $\mu_y$ denoted by $\mu_x \boxplus \mu_y.$  If at least one of the elements $x$ and $y$ is positive, then the distribution of $xy$ is called the free multiplicative convolution of $\mu_x$ and $\mu_y$, denoted by $\mu_x\boxtimes\mu_y$.
Moreover, for any integer $k,$ the free convolution power $\mu^{\boxplus k}$ is meaningful in the context of free probability theory. More generally, the range of $k$ can be extended to any real number, and $\mu^{\boxplus t}$ forms a semi-group for any $t\geq 1$ \cite{NS2006}. 

\vspace{2mm}

\noindent{\it Estimates for the support of free convolution power of given measure}--Suppose that $x \in \A$ is a self-adjoint element with distribution $\mu,$ and given a projection $p \in \A$ free from $x$ with $\varphi (p) = t \in (0, 1].$ Then the distribution of $t^{-1} p x p$ is the free convolution power $\mu^{\boxplus 1/t}$ \cite{NS96}.  Usually, It is highly non-trivial to study the support of $\mu^{\boxplus 1/t}$ (see \cite{BV95, B97, H12}). In this paper, we will use the following estimates due to Collins, Fukuda and Zhong:
\begin{prop}\label{prop:est-free-convolution-power}\cite[Lemma 2.3]{BM2015}. For any self-adjoint element $x \in \A$ such that $A \leq x \leq B$ with $A, B \in \mathbb{R},$ and denote $m(x)$ and $\sigma(x)^2$ by the mean and variance of $x$ respectively, we have for $k >1$
$${\rm supp} (\mu^{\boxplus k}) \subseteq [x_1(k), x_2 (k)],$$
where 
$$x_1(k)= A-2 \sigma(x) \sqrt{k-1}+ (k-1) m(x),$$
$$x_2(k)= B+2 \sigma(x) \sqrt{k-1}+ (k-1) m(x).$$
\end{prop}

\vspace{2mm}

\noindent {\it Strongly convergence in distribution}--Recall that a $C^{\ast}$-probability space is a $\ast$-probability space $(\A,\varphi)$ where $\A$ is an unital $C^{\ast}$-algebra. We have the following definition for the strongly convergence of noncommutative random variables 

\begin{defn}\cite{BC2014}.
For a sequence of $k$-tuple of random variables $({\bf x}_N)_{N \geq 1}= (x_N^{(1)}, \ldots, x_N^{(k)})_{N\geq 1}$ in $C^*$-probability spaces $(\A_N, \varphi_N),$ if the map
$$P \rightarrow \varphi_N \left[ P({\bf x}_N, {\bf x}_N^*)\right]$$
converges pointwisely for any polynomials $P$ in $2k$ noncommutative indeterminates, then we say $({\bf x}_N)_{N \geq 1}$ converges in distribution. And by strongly convergence in distribution, we mean convergence in distribution, and pointwise convergence of the map
$$P \rightarrow \left\| P({\bf x}_N, {\bf x}_N^*)\right\|.$$ 
\end{defn}

We remark that in Definition \ref{def:1}, if there exists $C^*$-probability space $(\A, \varphi)$ and a self-adjoint variable $x \in \A$ such that the distribution of $x$ in $\A$ is $\mu,$ then $X_N$ strongly converges in distribution to $x$ as $N \rightarrow \infty.$


\section{Strongly convergence of random matrices}


\noindent {\it Strongly convergence of (random) matrices}--We recall that  for any probability measure $\mu$ on the real line, its distribution function is defined by $F_{\mu}(z) : = \mu (( -\infty, z]).$ For a sequence of probability measures $\{\mu_N\}$, we say $\mu_N$ converges (weakly) to $\mu$ as $N \rightarrow \infty$, if $F_{\mu_N} (z) \rightarrow F_{\mu} (z)$ as $N \rightarrow \infty.$

For any Hermitian matrix $X_N \in \M_N(\mathbb{C})$, its empirical eigenvalue distribution is given by 
$$\mu_N = \frac{1}{N} \sum_{i=1}^{N} \delta_{\lambda_i},$$ 
where $\lambda_i$'s are the eigenvalues of $X_N.$ Hence we have
$${\rm tr}_{N} \left[ X_N^p \right] = \int_{\R} t^p d\mu_N (t), \; \text{for any integer} \; p\geq 1.$$
Motivated by \cite{CHN2016, CM2012}, we have the following definition:

\begin{defn}\label{def:1}
Let $X_N \in \M_N(\mathbb{C})$ be a (random) Hermitian matrix, for a given compactly supported probability measure $\mu,$ if almost surely 
\begin{enumerate}[{\rm(i)}]
\item $\mu_N$ converges to $\mu$ as $N \to \infty$;

\item The extremal eigenvalues of $X_N$ converge to the respective extrema of the support of $\mu$, i.e.,
$$\lim_{N \rightarrow \infty} \lambda_{min} (X_N) = \min \{z: z \in {\rm supp} (\mu)\}$$
$$\lim_{N \rightarrow \infty} \lambda_{max} (X_N) = \max \{z: z \in {\rm supp} (\mu)\},$$
\end{enumerate}
where $\lambda_{\min}$ (resp. $\lambda_{\max}$) is the smallest (resp. largest) eigenvalue. Then we can say that almost surely as $N \to \infty,$ $X_N$ strongly converges in distribution to $\mu$. 
\end{defn}

\begin{prop}\cite[Proposition 3.15-3.17]{NS2006}\label{prop:norm}
	Let $(\A,\varphi)$ be a $C^{\ast}$-probability space such that $\varphi$ is faithful. Let $x \in\A$ be a normal element (which means $xx^*= x^* x$) and $\mu$ be its $\ast$-distribution. Then the support of $\mu$ is equal to the spectrum of $x$, i.e.,
	\[ { \rm spec} [x]= { \rm supp} (\mu).\]
	Moreover, for any $x\in\A$ we have 
	\[ \| x \|=\lim_{k \to\infty}\varphi \left[ (x^*x)^{k} \right]^{\frac{1}{2k}}.\]
\end{prop}

\vspace{2mm}

\noindent {\it The asymptotic freeness of independent random matrices}--In this paper, the entries of $X_N$ are some random variables in a given probability space. And we additionally suppose that the distribution of $X_N$ is Haar unitary invariant. In random matrix theory, there are two important Haar unitary invariant ensembles. One is the Gaussian unitary ensemble (GUE), and the other is the complex Wishart ensemble.  We will focus on the latter one. For given parameter $\lambda,$ let $B_N$ be a $d \times N$ rectangle Gaussian random matrix. Assume that $\lambda = d/N$ as $d, N \to \infty.$ Write $W_N = B_N^* B_N$, then 
$W_N$ is called a $N \times N$ Wishart random matrix. 

In \cite{HT2003, MC2007, MM2004}, the strong convergence in distribution of GUE and Wishart ensembles are studied. Thus almost surely, the two ensembles embrace the conditions of Definition \ref{def:1}. For GUE, the strong limit fulfills the famous semicircular law, whose density is given by 
$$\mu_{SC}(t)= \frac{1}{2\pi} \sqrt{4-x^2} \un_{[-2,2]} (t) dt.$$  
And for the Wishart ensemble, the strong limit satisfies the Marcenko-Pastur law (with parameter $\lambda$), which is the free analog of the Poisson distribution with parameter $\lambda.$ The density is given by
$$\mu_{MP}(t)= \max\{1-\lambda, 0\} \delta_0 + \frac{\sqrt{(t-a) (b-t)}} {2\pi t} \cdot \un_{[a, b]} (t) dt,$$
where $a= (\sqrt{\lambda}-1)^2$ and $b=(\sqrt{\lambda}+1)^2.$

For independent random matrices $X_N$ and $Y_N \in \M_N(\mathbb{C})$, it is natural to ask whether its sum or product has a strong limit? In a series of works by Haagerup, Thorbj{\o}rnsen, Collins, Male, et al., this problem was properly addressed. We adapt the related results as the following proposition.

 \begin{prop}\label{prop:strong}\cite{HT2005, CM2012, BC2014, BM2015}.
For independent Hermitian random matrices $X_N$ and $Y_N,$ assume that:
\begin{enumerate}[{\rm (i)}]
\item almost surely as $N \to \infty$, $X_N$ and $Y_N$ strongly converges in distribution to $\mu$ and $\nu$ respectively;
\item at least one of $X_N$ and $Y_N$ is Haar unitary invariant, which means their laws are invariant under unitary conjugacy. 
\end{enumerate}
Then almost surely as $N \to \infty,$ $X_N + Y_N$ and $X_N Y_N$ strongly converges in distribution to $\mu \boxplus \nu$ and  $\mu\boxtimes\nu$ respectively. Especially, suppose that $X_N$ is Haar unitary invariant and $Y_N$ is a projection with ${\rm tr}_{N}(Y_N) =t$, then almost surely as $N \to \infty,$ $ Y_NX_N Y_N$ strongly converges in distribution to $\mu \boxtimes \left[ (1-t)\delta_0+ t \delta_1\right].$ 
 \end{prop}
  
 The above results reveal a deep relation between the free probability theory and random matrix theory. Namely, if we consider $X_N$ and $Y_N$ as the random variables in the $*$-probability space $( \M_N(\mathbb{C}), {\rm tr}_{N}),$ then almost surely as $N \to \infty$, $X_N$ and $Y_N$ are freely independent. This phenomenon is called (strong) asymptotic freeness of random matrices, which was firstly found by Voiculescu, and the strong convergence was initially studied by Haagerup and Thorbj{\o}rnsen. 
 
In their seminal work, Haagerup and Thorbj{\o}rnsen proved the strong asymptotic freeness of independent GUE matrices \cite{HT2005}. Their result was extended by Male \cite{CM2012} to independent GUE matrices plus an extra family of independent matrices with strong limiting distribution, and later on, Collins and Male proved similar results for independent unitary Haar matrices \cite{BC2014}. 


\section{Strongly convergence of partial transposition of block random matrices}

In this section, we consider the following (block) Hermitian matrix 
\begin{equation}\label{eq:block-matrix}
X_{nN} = \sum_{i,j=1}^n E_{ij} \otimes X_{ij} \in \M_n (\mathbb{C}) \otimes \M_N(\mathbb{C}),
\end{equation}
where $\{E_{ij}\}_{i,j=1}^n$ is the unit basis of $\M_n(\mathbb{C}).$ The partial transposition of $X_{nN}$ is given by
\begin{equation}
X_{nN}^\Gamma : = \sum_{i,j=1}^n E_{ij} \otimes X_{ji}.
\end{equation}
Suppose that there exists a probability measure $\mu$ on $\mathbb{R}$ such that $X_{nN}$ strongly converges to $\mu$ as $N \rightarrow \infty.$ A natural question is that how about the convergence of 
$X_{nN}^\Gamma?$

We will consider the problem in the following compressed  $*$-probability space of $(\mathcal{A}, \varphi)$. 
\begin{defn}\label{def:matrix units}\cite{NS2006}.
Let $(\mathcal{A}, \varphi)$ be a $*$-probability space, we call the elements $\{e_{ij}\}_{i,j=1}^n \subseteq \mathcal{A}$ a family of matrix units of $\mathcal{A}$, if the following conditions hold:
\begin{enumerate}[{\rm(i)}]
\item $e_{ij}^* = e_{ji}$ and $e_{ij} e_{kl} = \delta_{j,k} e_{il}$ for all $i,j,k,l=1, \ldots, n;$

\item $\sum_{i=1}^n e_{ii}= e,$ where $e$ is the unit of $\mathcal{A};$

\item $\varphi(e_{ij})=0$ whenever $i \neq j$ and $\varphi(e_{11}) = \cdots = \varphi(e_{nn}) = 1/n.$
\end{enumerate}
We denote by $(\tilde{\mathcal{A}}, \psi)$ the compression of $(\mathcal{A}, \varphi)$ by $e_{11},$ i.e., $\tilde{\mathcal{A}}: = e_{11} \mathcal{A} e_{11}$ and $\psi: = n \cdot \varphi|_{\tilde{\mathcal{A}}}.$
It is known that $(\tilde{\A},\psi)$ is also a $\ast$-probability space.
\end{defn}

Let $x$ be a self-adjoint element of $\mathcal{A}$ with distribution $\mu$, and it is free from $\{e_{ij}\}_{i,j=1}^n$. We consider the following family of elements 
$$x_{ij}: = e_{1i} x e_{j1} \in \tilde{\mathcal{A}}, \; i,j=1, \ldots, n.$$ 
Let
\begin{equation}\label{eq:tilde{x}}
\tilde{x} = \sum_{i,j=1}^n E_{ij} \otimes x_{ij} 
\end{equation}
It is known \cite[Theorem 3.2]{CHN2016} that the distribution of $\tilde{x}$ in the $*$-probability space $(\M_n(\mathbb{C}) \otimes \tilde{\mathcal{A}}, {\rm tr}_n \otimes \psi)$ coincides with the distribution of $x$ in $(\mathcal{A}, \varphi)$. Let
\begin{equation}\label{eq:transpose-x}
 \tilde{x}^\Gamma = \sum_{i,j=1}^n E_{ij} \otimes x_{ji}  \in \M_n(\mathbb{C}) \otimes \tilde{\mathcal{A}}.
\end{equation}
We denote $\mu^\Gamma$ by the distribution of $\tilde{x}^\Gamma$ in  $(\M_n(\mathbb{C}) \otimes \tilde{\mathcal{A}}, {\rm tr}_n \otimes \psi).$

\begin{rem}
There is a natural model for $\mathcal{A}$ and $\tilde{\mathcal{A}}$ in Definition \ref{def:matrix units} \cite{CHN2016}. Recall that for a given compactly supported measure $\mu,$ let $x$ be a self-adjoint element in the $*$-probability space $(L_\infty(\mathbb{R}, \mu), \mathbb{E}).$ Let $\mathcal{A} = \M_n(\mathbb{C}) \ast L_\infty(\mathbb{R})$ be the free product of $\M_n(\mathbb{C})$ and $L_\infty(\mathbb{R}),$ then $\{E_{ij}\}_{i,j=1}^n$ is a family of matrix units of $\mathcal{A},$ and obviously $x$ is free from $\{E_{ij}\}_{i,j=1}^n$ in $\mathcal{A}$ (see \cite{NS2006}). 
\end{rem}

\begin{lem}\label{lem:wc}
\begin{enumerate}[{\rm (a)}]
\item For any element $y \in \M_n(\mathbb{C}) \otimes \tilde{\A}$ we have
	\begin{equation}
	\left\| y \right\|_{\M_n \otimes \tilde{\A}} = \left\| y \right\|_{\M_n \otimes \A}.
	\end{equation}
\item 
Let $E_{11} \M_n(\mathbb{C}) E_{11} = \{E_{11} X E_{11}: X \in \M_n(\mathbb{C})\},$ then $(E_{11} \M_n(\mathbb{C}) E_{11}, n \cdot {\rm tr}_n)$ is a $C^*$-probability space with a faithful state
$n \cdot {\rm tr}_n.$ For any matrix $Y \in \M_n(\mathbb{C})\otimes(E_{11}\M_n(\mathbb{C})E_{11})\otimes \M_N(\mathbb{C})$ we have
\begin{equation}
	\left\| Y \right\|_{\M_n \otimes(E_{11}\M_n E_{11})\otimes \M_N} = \left\| Y \right\|_{\M_n \otimes \M_n \otimes \M_N}.
	\end{equation}
	\end{enumerate}
\end{lem}

\begin{proof}
We only prove (a), and the proof for (b) is similar. By Proposition \ref{prop:norm}, we have 
\begin{align*}
	 \left\| y \right\|_{\M_n \otimes \tilde{\A}} & =\lim_{k\to\infty} \left(  {\rm tr}_n \otimes \psi \left[ (y^*y)^{k}  \right] \right)^{\frac{1}{2k}}\\
	& = \lim_{k\to\infty} n^{\frac{1}{2k}} \cdot \left( {\rm tr}_n \otimes \varphi \left[ (y^*y)^{k}  \right] \right)^{\frac{1}{2k}} = \lim_{k\to\infty}  \left( {\rm tr}_n \otimes \varphi \left[ (y^*y)^{k}  \right] \right)^{\frac{1}{2k}}\\
	& = \left\| y \right\|_{\M_n \otimes \A}.
			\end{align*}
\end{proof}

The following lemma is based on Haagerup and Thorbj{\o}rnsen \cite[Theorem 9.1]{HT2005}. 

 \begin{lem}\label{Msc}\cite{HT2005}.
	For integer $N \geq1$, let ${\bf X}_{nN}= \{ X_{nN}^{(1)}, \ldots, X_{nN}^{(r)} \}$ be an $r$-tuple of random matrices in the $C^*$-probability space $(\M_{nN} (\mathbb{C}), {\rm tr}_{nN})$, and let ${\bf x}= \{x_1, \ldots, x_r\}$ be an $r$-tuple of random variables in the $C^{\ast}$-probability space $(\mathcal{A},\varphi)$. Assume that: almost surely, 
	$${\bf X}_{nN} \xrightarrow{N \to \infty} {\bf x}, \; \text{ strong convergence in distribution}.$$
	
	Then for any polynomial $P$ in $r$ noncommutative indeterminates with coefficients in $\M_{n}(\mathbb{C})$, almost surely, we have 
	 \[\lim_{ N \to\infty} \left\| P(X_{nN}^{(1)},\cdots,X_{nN}^{(r)}) \right\|_{\M_{n}\otimes \M_{nN}}=\left\|P(x_1,\cdots,x_{r})\right\|_{\M_{n}\otimes\mathcal{A} }. \] 
\end{lem}

\begin{thm}\label{thm:PT}
For Hermitian matrix $X_{nN}$ given in Equation \eqref{eq:block-matrix}, assume that
\begin{enumerate}[{\rm (i)}]
\item  $X_{nN}$ is Haar unitary invariant;
\item  almost surely as $N \to \infty$, $X_{nN}$ strongly converges in distribution to a probability measure $\mu$.
\end{enumerate}
Then almost surely as $N \to \infty,$ $X_{nN}^\Gamma$ strongly converges in distribution to $\mu^\Gamma.$ 
\end{thm}

\begin{proof}
We note that in \cite [Proposition 4.1]{ANV16}, the authors showed the convergence of $X_{nN}^\Gamma,$ which was updated to the strongly convergence in 
\cite[Theorem 6]{IN2018}. We present an alternative proof here.

Let us consider the following $C^*$-probability space $\big(\M_n(\mathbb{C})\otimes(E_{11}\M_n(\mathbb{C})E_{11})\otimes \M_N(\mathbb{C}), {\rm tr}_n \otimes (n\cdot {\rm tr}_n) \otimes {\rm tr}_N \big).$ Let
\begin{equation*}
\begin{split} 
C_N & :=\sum_{i,j=1}^{n}E_{ij}\otimes((E_{1j}\otimes \un_N)X_{nN} (E_{i1}\otimes \un_N))\\
& = \sum_{i,j=1}^n E_{ij} \otimes E_{11} \otimes X_{ji}.
\end{split}
\end{equation*}
An elementary computation shows that 
\begin{equation}
\tr_{nN} \left[ \left(X_{nN}^\Gamma \right)^p \right] = {\rm tr}_n \otimes (n\cdot {\rm tr}_n) \otimes {\rm tr}_N \left[ C_N^p \right]
\end{equation}
for any integer $p\geq 1.$

Suppose that there exists $C^*$-probability space $(\A, \varphi)$ and a self-adjoint element $x \in \A$ such that the distribution of $x$ is $\mu.$ Suppose ${\bf e}= \{e_{ij}\}_{i,j=1}^{n}\subseteq (\mathcal{A}, \varphi)$ be the family of matrix units of $\A,$ which is free from $x$ in $\A.$ 
We note that $(\A, \varphi), x,$ and  ${\bf e}$ can be explicitly constructed (see \cite{CHN2016}).
Let ${\bf E}_N= \{E_{ij} \otimes \un_N\}_{i,j=1}^n$ be an $n^2$-tuple of matrices in the $C^*$-probability space $(\M_{nN} (\mathbb{C}),  {\rm tr}_{nN}),$Thus by \cite{BC2014}, almost surely as $N \rightarrow \infty$, $\{X_{nN}, {\bf E}_N\}$
strongly converges to $\{x, {\bf e}\}.$ Hence for any polynomial $P$ with $n^2$ noncommutative indeterminates, almost surely we have
\begin{enumerate}[{\rm (a)}]
\item ${\rm tr}_{nN} \left[  P\left( \left\{ E_{11} \otimes X_{ij} \right\}_{i,j=1}^n  \right) \right] \stackrel{N \rightarrow \infty} {\longrightarrow} \varphi \left[  P\left( \left\{ x_{ij} \right\}_{i,j=1}^n \right) \right];$
\item $\left\|  P\left( \left\{ E_{11} \otimes X_{ij} \right\}_{i,j=1}^n  \right) \right\|_{\M_{nN}(\mathbb{C})} \stackrel{N \rightarrow \infty} {\longrightarrow} \left\| P\left( \left\{ x_{ij} \right\}_{i,j=1}^n \right) \right\|_{\A}.$
\end{enumerate}

By (a) we have 
\begin{equation*}
\begin{split}
{\rm tr}_n \otimes (n\cdot {\rm tr}_n) \otimes {\rm tr}_N \left[ C_N^p \right] & = \sum_{i_1, j_1, \ldots, i_p, j_p=1}^n {\rm tr}_n \left[ E_{i_1 j_1} \cdots E_{i_p j_p}\right] \cdot n \cdot {\rm tr_{nN}} \left[ E_{11} \otimes \left( X_{j_1i_1} \cdots X_{j_p i_p} \right)\right]\\
& \stackrel{N \rightarrow \infty} {\longrightarrow}  \sum_{i_1, j_1, \ldots, i_p, j_p=1}^n {\rm tr}_n \left[ E_{i_1 j_1} \cdots E_{i_p j_p}\right] \cdot n \cdot \varphi \left( x_{j_1i_1} \cdots x_{j_p i_p} \right)\\
& = {\rm tr_n} \otimes \psi \left[ \left(\tilde{x}^\Gamma\right)^p \right].
\end{split} 
\end{equation*}
Denote $\mu_{N}^\Gamma$ by the empirical eigenvalue distribution of $X_{nN}^\Gamma,$ then we show that $\mu_N^\Gamma$ converges to $\mu^\Gamma.$ 

Moreover, by combining (b), Lemma \ref{lem:wc}, and Lemma \ref{Msc}, almost surely we have
\begin{equation*} 
\begin{split}
\lim_{N \rightarrow \infty} \left\| C_N \right\|_{\M_n \otimes (E_{11} \M_n E_{11}) \otimes \M_N} & =
\lim_{N \rightarrow \infty} \left\| C_N \right\|_{\M_n \otimes \M_{nN}} \\
& = \lim_{N \rightarrow \infty} \left\| \sum_{i,j=1}^n E_{ij} \otimes E_{11} \otimes X_{ji}\right\|_{\M_n \otimes \M_{nN}}\\
& =  \left\| \sum_{i,j=1}^n E_{ij} \otimes x_{ji}\right\|_{\M_n \otimes \A} = \left\| \tilde{x}^\Gamma \right\|_{\M_n \otimes \A},
\end{split}
\end{equation*}
which concludes our proof. 
\end{proof}


\section{Limit distribution of $X_{nN}^\Gamma$}

\begin{defn}\cite{NS2006}\label{ccw}
Given a partition $\pi\in NC(n)$ and an $n$-tuple
of double-indices $(i_1 j_1, i_2 j_2, \ldots, i_nj_n)$, we
say that $\pi$ couples in a cyclic way (c.c.w., for short) the indices
$(i_1 j_1, i_2 j_2, \ldots, i_nj_n)$ if we have for each block $(l_1 < l_2 < \cdots < l_s) \in \pi$ that $j_{l_k} = i_{l_{k+1}}$ for all $k = 1, . . . , s$ (where we put $l_{s+1} := l_1$).
\end{defn}
\begin{notation}
	For any noncommutative probability space $(\mathcal{A},\varphi)$, we use $(\kappa_\pi^\mathcal{A})_{\pi\in NC(n)}$ to denote the free cumulants functionals on $(\mathcal{A},\varphi)$ which are given by Equation \eqref{eq:moment-cumulant}.
\end{notation}

For the completeness of the proof, we rewrite the theorem \cite[Theorem 14.18]{NS2006} as the following lemma. 
 
\begin{lem}\label{lem:1}\cite[Theorem 14.18]{NS2006}
Let $(\mathcal{A},\varphi)$ be a $*$-probability space and $\{x_{1}, x_{2}, \ldots, x_r\}$ be a sequence of random variables in $\mathcal{A}.$ Let $\{e_{ij}\}_{i,j=1}^n \subseteq \mathcal{A}$ be a family of matrix units such that
	\[\varphi(e_{ij})=\frac{1}{n} \delta_{ij}, \ \  i, j =1,\ldots,n,\]
	and $\{x_{1}, x_{2}, \ldots, x_r\}$ is freely independent from $\{e_{ij}\}_{i,j=1}^n$. Denote $x_{ij}^{(s)}:=e_{1i}x_{s}e_{j1}$ and $\lambda:=\varphi(e_{11})=1/n,$ 
then for $p \geq 1$  we have
	\[\kappa_p^{\tilde{\mathcal{A}}} \left[ x_{i_{1}j_{1}}^{(s_{1})},\ldots, x_{i_{p}j_{p}}^{(s_{p})} \right] =\begin{cases}
	1/\lambda \cdot \kappa_{p}^{\mathcal{A}} \left[ \lambda x_{s_1},\ldots,\lambda x_{s_p} \right]  & \; j_{k}=i_{k+1}, k=1,\ldots, p, \\
	0 & \; \text{otherwise},
	\end{cases}\]
where $s_{1},\ldots, s_{p} \in \{1, \ldots, r\}$, $i_{1}, j_{1}, \ldots, i_{p}, j_{p} \in \{1, \ldots, n\}.$
\end{lem}

\begin{thm}\label{thm:1}
Let $(\mathcal{A},\varphi)$ be a $*$-probability space and $x$ be a self-adjoint random variable in $\mathcal{A}.$ Let $\{e_{ij}\}_{i,j=1}^n \subseteq \mathcal{A}$ be a family of matrix units. Recall $(\tilde{A}, \psi)$ is the compression of $(\mathcal{A}, \varphi)$ by $e_{11},$ and $\tilde{x}^\Gamma$ is given by Equation \eqref{eq:transpose-x}.
Then the free cumulants (with respect to $(\mathcal{M}_{n}(\mathbb{C})\otimes \tilde{\mathcal{A}}, {\rm tr}_n \otimes \psi)$) of $\tilde{x}^\Gamma$ can be computed by the 
 free cumulants (with respect to $(\mathcal{A}, \varphi)$) of $x$ as follows: 
 \begin{equation}
\kappa_p \left[ \tilde{x}^\Gamma, \tilde{x}^\Gamma, \ldots, \tilde{x}^\Gamma \right]= c(n,p) \cdot \kappa_{p} \left[ \frac{x}{n},  \frac{x}{n}, \ldots,  \frac{x}{n} \right],
\end{equation}
where $c(n,p)$ is a constant satisfies 
\begin{equation*}
c(n,p)=
\begin{cases}
	n  &  \; \text{for odd} \; p, \\
	n^2  & \; \text{for even} \; p.
	\end{cases}
\end{equation*}
\end{thm}

\begin{proof}
Let us consider the following $\ast$-moments of $\tilde{x}^\Gamma$ in $(\mathcal{M}_{n}(\mathbb{C})\otimes \tilde{\mathcal{A}}, {\rm tr}_n \otimes \psi)$.	Denote $x_{ij}:=e_{1i}xe_{j1},$ for any integer $p\geq 1,$ by Equation \eqref{eq:moment-cumulant} we have 
	\begin{equation*}
	\begin{split}
	 {\rm tr_n} \otimes \psi \left[ \left( \tilde{x}^\Gamma \right)^p\right] & = {\rm tr_n} \otimes \psi \left[ \sum_{i_{1},j_{1},\ldots, i_{p},j_{p}=1}^{n}E_{i_{1}j_{1}}E_{i_2j_2}\cdots E_{i_{p}j_{p}}\otimes x_{j_1i_1} x_{j_2i_2}\cdots x_{j_p i_p} \right] \\
	& = \frac{1}{n} \sum_{i_{1},\ldots, i_{p}=1}^{n} \psi \left[ x_{i_2i_1} x_{i_3i_2} \cdots  x_{i_1i_p} \right] \\
	& = \frac{1}{n} \sum_{i_{1},\ldots, i_{p}=1}^{n} \sum_{\pi\in NC(p)} \kappa_\pi^{\tilde{\mathcal{A}}} \left[ x_{i_2i_1}, x_{i_3 i_2} \ldots, x_{i_1i_p} \right]\\
	& = \frac{1}{n} \sum_{\pi\in NC(p)} \sum_{i_{1},\ldots, i_{p}=1}^{n} \prod_{V \in \pi}\kappa^{\tilde{\mathcal{A}}}(\mathbf{1}_{|V|}) \left[ x_{i_2i_1}, x_{i_3i_2}, \ldots, x_{i_1i_p}\right].
	\end{split}
	\end{equation*}
				
For any block  $V=\{l_1,\ldots, l_s\} \in \pi$, by Lemma \ref{lem:1} the term $\kappa^{\tilde{\mathcal{A}}}(\mathbf{1}_{|V|}) \left[ x_{i_2i_1}, x_{i_3i_2}, \ldots, x_{i_1i_p}\right]$
does not vanish only if
\begin{equation}\label{eq:condition-1}
i_{l_1} = i_{l_2+1}, i_{l_2} = i_{l_3+1}, \ldots, i_{l_s} = i_{l_1+1}.
\end{equation}
It follows that the term $\kappa_\pi^{\tilde{\mathcal{A}}} \left[ x_{i_2i_1}, x_{i_3 i_2} \ldots, x_{i_1i_p} \right]$
does not vanish only if $\pi$ c.c.w. (recall Definition \ref{ccw}) the $p$-tuple of double-indices $(i_{2}i_{1},i_{3}i_{2},\ldots,i_{1}i_{p})$. 
Thus the term $\kappa_{\pi}^{\tilde{\mathcal{A}}} \left[ x_{i_2i_1}, \ldots, x_{i_1i_p} \right] $ can be computed as follows:
	\begin{equation}
	\begin{split}  
	\kappa_{\pi}^{\tilde{\mathcal{A}}} \left[ x_{i_2i_1}, \ldots, x_{i_1i_p} \right] & =
	\begin{cases}
	\prod\limits_{V \in \pi} n \cdot \kappa^{\mathcal{A}} (\mathbf{1}_{|V|}) \left[ \frac{x}{n}, \frac{x}{n}, \ldots, \frac{x}{n} \right] &\pi\  \text{c.c.w.}\  (i_{2}i_{1},i_{3}i_{2},\ldots,i_{1}i_{p}), \notag\\
	0 & \text{otherwise}.
	\end{cases}\notag\\ 
	& =  
	\begin{cases}
	n^{|\pi|} \cdot \kappa_{\pi}^{\mathcal{A}} \left[ \frac{x}{n}, \frac{x}{n}, \ldots, \frac{x}{n} \right] & \pi\  \text{c.c.w.}\  (i_{2}i_{1},i_{3}i_{2},\ldots,i_{1}i_{p}), \notag\\
	0 & \text{otherwise}.
	\end{cases}
	\end{split}
	\end{equation}
Hence, it remains to count the number of the sequence $\{i_{1},\ldots, i_{p}\}$ such that $\pi$ c.c.w. the tuple of double-indices $(i_{2}i_{1},i_{3}i_{2},\ldots,i_{1}i_{p})$. We denote this number by $N (\pi)$.      
				
We firstly let $\pi=\mathbf{1}_{p}$, and we denote that $i_{p+1}: =i_{1}$ and $i_{p+2}:= i_{2}$. 
Now we will use the following graphs to present how the partition $\mathbf{1}_{p}$ c.c.w. the $p$-tuple of double indices $(i_{2}i_{1},i_{3}i_{2},\ldots,i_{1}i_{p})$. In the graph, indices will be connected if they are given by the same value. 
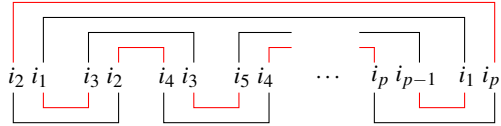
\begin{figure}[H]
\centering
\begin{tikzpicture}	
	\node (P1) at (0,0) {$i_2\ i_1$};
	\node (P2) at (1,0) {$i_3\ i_2$};
	\node (P3) at (2,0) {$i_4\ i_3$};
	\node (P4) at (3,0) {$i_5\ i_4$};
	\node (P5) at (4,0) {$\cdots$};
	\node (P6) at (5,0) {$i_p\ i_{p-1}$};
	\node (P7) at (6,0) {$i_1\ i_p$};
	\draw (0.2,0.2)--(0.2,0.8)--(5.8,0.8)--(5.8,0.2);
	\draw (-0.2,-0.2)--(-0.2,-0.6)--(1.2,-0.6)--(1.2,-0.2);
	\draw (0.8,0.2)--(0.8,0.6)--(2.2,0.6)--(2.2,0.2);
	\draw (1.8,-0.2)--(1.8,-0.6)--(3.2,-0.6)--(3.2,-0.2);
	\draw (2.8,0.2)--(2.8,0.6)--(3.5,0.6);
	\draw (5.2,0.2)--(5.2,0.6)--(4.4,0.6);
	\draw (4.6,-0.2)--(4.6,-0.6)--(6.2,-0.6)--(6.2,-0.2);
	\draw[red] (0.2,-0.2)--(0.2,-0.4)--(0.8,-0.4)--(0.8,-0.2);
	\draw[red] (2.2,-0.2)--(2.2,-0.4)--(2.8,-0.4)--(2.8,-0.2);
	\draw[red] (5.2,-0.2)--(5.2,-0.4)--(5.8,-0.4)--(5.8,-0.2);
	\draw[red] (1.2,0.2)--(1.2,0.4)--(1.8,0.4)--(1.8,0.2);
	\draw[red] (3.2,0.2)--(3.2,0.4)--(3.5,0.4);
	\draw[red] (4.6,0.2)--(4.6,0.4)--(4.4,0.4);
	\draw[red] (-0.2,0.2)--(-0.2,1)--(6.2,1)--(6.2,0.2);

\end{tikzpicture}
\caption{$p$ is even}
\end{figure}
\begin{figure}[H]
\centering
\begin{tikzpicture}
	\node (P1) at (0,0) {$i_2\ i_1$};
	\node (P2) at (1,0) {$i_3\ i_2$};
	\node (P3) at (2,0) {$i_4\ i_3$};
	\node (P4) at (3,0) {$i_5\ i_4$};
	\node (P5) at (4,0) {$\cdots$};
	\node (P6) at (5,0) {$i_{p-1}\ i_{p-2}$};
	\node (P7) at (6.5,0) {$i_{p}\ i_{p-1}$};
	\node (P8) at (7.5,0) {$i_1\ i_{p}$};
	\draw (0.2,0.2)--(0.2,0.8)--(7.35,0.8)--(7.35,0.7);
	\draw (7.35,0.5)--(7.35,0.2);
	\draw (7.35,0.5) arc (-90:90:0.1);
	\draw (-0.2,-0.2)--(-0.2,-0.6)--(1.2,-0.6)--(1.2,-0.2);
	\draw (0.8,0.2)--(0.8,0.6)--(2.2,0.6)--(2.2,0.2);
	\draw (1.8,-0.2)--(1.8,-0.6)--(3.2,-0.6)--(3.2,-0.2);
	\draw (2.8,0.2)--(2.8,0.6)--(3.5,0.6);
	\draw (5.2,0.2)--(5.2,0.6)--(4.4,0.6);
	\draw (4.6,-0.2)--(4.6,-0.6)--(6.5,-0.6)--(6.5,-0.2);
	\draw (6.1,0.2)--(6.1,0.6)--(7.6,0.6)--(7.6,0.2);
	\draw[red] (0.2,-0.2)--(0.2,-0.4)--(0.8,-0.4)--(0.8,-0.2);
	\draw[red] (2.2,-0.2)--(2.2,-0.4)--(2.8,-0.4)--(2.8,-0.2);
	\draw[red] (5.2,-0.2)--(5.2,-0.4)--(6.1,-0.4)--(6.1,-0.2);
	\draw[red] (1.2,0.2)--(1.2,0.4)--(1.8,0.4)--(1.8,0.2);
	\draw[red] (3.2,0.2)--(3.2,0.4)--(3.5,0.4);
	\draw[red] (4.6,0.2)--(4.6,0.4)--(4.4,0.4);
	\draw[red] (-0.2,0.2)--(-0.2,1)--(7.65,1)--(7.65,0.2);
	\draw[red] (6.5,0.2)--(6.5,0.4)--(7.3,0.4)--(7.3,0.2);
\end{tikzpicture}
\caption{$p$ is odd}
\end{figure}
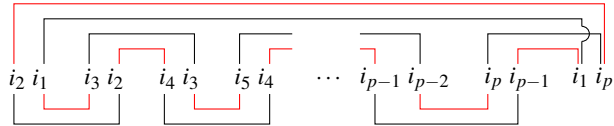

We can conclude that 
\begin{equation*}
N(\mathbf{1}_{p}) = n^{\# (\text{circles})}.
\end{equation*}
For odd $p$, due to the conditions \eqref{eq:condition-1}, we have 
$i_{1}=i_{3}=\cdots=i_{p}=i_{p+2}=i_{2}=i_{4}=\cdots=i_{p-1},$ thus $N (\mathbf{1}_{p})=n$ (see Figure 1).
For even $p$, we have $i_{1}=i_{3}=\cdots=i_{p-1}=i_{p+1},\ i_{2}=i_{4}=\cdots=i_{p}=i_{p+2}$,
thus $N (\mathbf{1}_{p}) =n^{2}$ (see Figure 2).


		
Now for general $\pi < \mathbf{1}_{p}$ in $NC(p)$, let $V=\{k+1,k+2,\ldots,k+q\}$ be a block of $\pi$ with $k \geq 0$ and $1\leq q, k+q\leq p$. For simplicity we denote $k+q+1= [k+q+1](\text{mod} \; p)$.
Suppose that $\pi|_{V}=\mathbf{1}_{q}$ c.c.w. the tuple $(i_{k+2}i_{k+1},i_{k+3}i_{k+2},\ldots,i_{k+q+1}i_{k+q}),$  then by the previous discussion we have 
\begin{align}\label{interval-block}
	\begin{cases}
	i_{k+1}=i_{k+2}=\cdots=i_{k+q}=i_{k+q+1} & \text{ for odd $q$,}\\
	i_{k+1}=i_{k+3}=\cdots=i_{k+q+1},\ i_{k+2}=i_{k+4}=\cdots=i_{k+q} &  \ \text{for even $q$.}
	\end{cases}
\end{align}
Let $\hat{i}_{k+1},\ldots,\hat{i}_{k+q} \in \{1,\ldots,n\}$ be some indices, then by Definition \ref{ccw} we have the following 
\begin{equation}\label{eq:equivalent-1}
\begin{split}
\pi \; \text{c.c.w.} \; (i_{2}i_{1},i_{3}i_{2},\ldots, i_{1}i_{p}) & \Longleftrightarrow \left\{
\begin{array}{l}
\pi \setminus V \; \text{c.c.w.} \;  (i_{2}i_{1},\ldots,i_{k+1}i_{k},i_{k+q+2}i_{k+q+1},\ldots,i_{1}i_{p});\\
\mathbf{1}_{|V|} \; \text{c.c.w.} \; (i_{k+2}i_{k+1},i_{k+3}i_{k+2},\ldots, i_{k+q+1}i_{k+q}).
\end{array} \right. \\
& \Longleftrightarrow \left\{
\begin{array}{l}
	\pi \setminus V \; \text{c.c.w.} \;  (i_{2}i_{1},\ldots,i_{k+1}i_{k},i_{k+q+2}i_{k+q+1},\ldots,i_{1}i_{p});\\
	\mathbf{1}_{|V|} \; \text{c.c.w.} \; (\hat{i}_{k+2} \hat{i}_{k+1}, \hat{i}_{k+3} \hat{i}_{k+2},\ldots, \hat{i}_{k+q+1} \hat{i}_{k+q}); \\
	i_{k+1}=\hat{i}_{k+1}, i_{k+2}=\hat{i}_{k+2},\ldots, i_{k+q+1}=\hat{i}_{k+q+1}.
\end{array} \right. 
\end{split}
\end{equation}
By \eqref{interval-block} we have $i_{k+1} = i_{k+q+1}$, then it follows that 
\begin{equation}\label{eq:equivalent-2}
\begin{split}
\pi \; \text{c.c.w.} \; (i_{2}i_{1},i_{3}i_{2},\ldots, i_{1}i_{p}) & \Longleftrightarrow \left\{
\begin{array}{l}
	\pi \setminus V \; \text{c.c.w.} \;  (i_{2}i_{1},\ldots,i_{k+1}i_{k},i_{k+q+2}i_{k+1},\ldots,i_{1}i_{p});\\
	\mathbf{1}_{|V|} \; \text{c.c.w.} \; (i_{k+2} \hat{i}_{k+1}, i_{k+3} i_{k+2},\ldots, \hat{i}_{k+1} i_{k+q}); \\
	i_{k+1}=\hat{i}_{k+1}.
\end{array} \right.
\end{split}
\end{equation}

Denote $N' (\pi\setminus V)$ by the number of sequence $(i_1,\ldots,i_{k+1},i_{k+q+2},\ldots,i_p)$ such that $\pi \setminus V$ c.c.w. the tuple of double-indices $(i_{2}i_{1},\ldots,i_{k+1}i_{k},i_{k+q+2}i_{k+1},\ldots,i_{1}i_{p})$, and $N'' (\mathbf{1}_{|V|})$ by the number of sequence $(\hat{i}_{k+1},i_{k+2},\ldots,i_{k+q})$ such that $\mathbf{1}_{|V|}$ c.c.w. the tuple of double-indices $(i_{k+2}\hat{i}_{k+1},i_{k+3}i_{k+2},\ldots,\hat{i}_{k+1}i_{k+q})$. For a fix $s \in \{1,2,\ldots,n\}$, we can deduce that
\begin{align*}
&N' (\pi \setminus V) = n \cdot \#\bigg\{\begin{pmatrix}
	i_1,\ldots,i_{k+1},\\ i_{k+q+2},\ldots, i_p
\end{pmatrix}\bigg|\begin{matrix}
\pi \setminus V \; \text{c.c.w.} \; (i_{2}i_{1},\ldots,i_{k+1}i_{k},i_{k+q+2}i_{k+1},\ldots,i_{1}i_{p}) \\ \text{and} \; i_{k+1}=s \\\end{matrix}\bigg\},
\end{align*}
and
\begin{align*}
&N'' (\mathbf{1}_{|V|})= n \cdot \#\bigg\{(\hat{i}_{k+1},i_{k+2},\ldots,i_{k+q}) \bigg|\begin{matrix}
 \mathbf{1}_{|V|} \; \text{c.c.w.} \; (i_{k+2}\hat{i}_{k+1},i_{k+3}i_{k+2},\ldots,\hat{i}_{k+1}i_{k+q})\\ \text{and} \; \hat{i}_{k+1}=s 
\end{matrix}\bigg\}.
\end{align*}
It follows from Equation \eqref{eq:equivalent-2} that 
	\[ N (\pi) = n \cdot \frac{N' (\pi\setminus V)}{n} \cdot \frac{N'' (\mathbf{1}_{|V|})}{n},\]
where the factor $n$ is due to $s$ has $n$ choices in $\{1,2,\ldots,n\}$.
Hence,  
	\begin{equation*}
	N (\pi) = \begin{cases}
	N' (\pi \setminus V) & \text{for odd $q$,}\\
	n \cdot N' (\pi\setminus V) & \text{for even $q$.}
	\end{cases}
        \end{equation*}
Therefore by induction we have 	    
	\[N (\pi)=n^{\alpha +1},\; \text{where} \; \alpha=\sum_{V \in\pi, |V| \; \text{is even}}1.\] 

Finally, we have 				
	\begin{align*}
  {\rm tr_n} \otimes \psi \left[ \left( \tilde{x}^\Gamma \right)^p\right] 
	& = \frac{1}{n} \sum_{i_{1},\ldots, i_{p}=1}^{n} \sum_{\pi\in NC(p)} \kappa_\pi^{\tilde{\mathcal{A}}} \left[ x_{i_2i_1}, x_{i_3 i_2} \ldots, x_{i_1i_p} \right] \notag\\
	& = \frac{1}{n}\sum_{\pi\in NC(p)}\sum_{\substack{i_{1},\ldots, i_{p}=1\\\pi\ \text{c.c.w.}\ (i_{2}i_{1},i_{3}i_{2},\ldots,i_{1}i_{p})}}^{n} n^{|\pi|} \cdot \kappa_{\pi}^{\mathcal{A}} \left[ \frac{x}{n},\frac{x}{n}, \ldots, \frac{x}{n} \right]\notag\\
	& = \frac{1}{n}\sum_{\pi\in NC(p)} n^{\alpha+|\pi|+1} \cdot \kappa_{\pi}^{\mathcal{A}} \left[ \frac{x}{n}, \frac{x}{n}, \ldots, \frac{x}{n} \right].
	\end{align*}
Note that the function
\begin{equation*}
\pi \rightarrow n^{\alpha+|\pi|} \cdot \kappa_{\pi}^{\mathcal{A}} \left[ \frac{x}{n}, \frac{x}{n}, \ldots, \frac{x}{n} \right]
\end{equation*}
is multiplicative, and by recalling the following moment-cumulant formula (see Equation \eqref{eq:moment-cumulant})
\begin{equation*}
{\rm tr_n} \otimes \psi \left[ \left( \tilde{x}^\Gamma \right)^p\right]  = \sum_{\pi\in NC(p)} \kappa_\pi \left[ \tilde{x}^\Gamma,  \tilde{x}^\Gamma,\ldots,  \tilde{x}^\Gamma \right],
\end{equation*}
	we can deduce our proof.
	\end{proof}

\begin{rem}
Suppose that $x$ is not a self-adjoint random variable in $(\mathcal{A}, \varphi),$ for any integer $p\geq 1$, let $\epsilon_1, \epsilon_2, \ldots, \epsilon_p \in \{1, *\}$, then by similar argument we have
\begin{equation}
\kappa_p \left[ \left(\tilde{x}^\Gamma\right)^{\epsilon_1}, \left(\tilde{x}^\Gamma\right)^{\epsilon_2}, \ldots, \left(\tilde{x}^\Gamma\right)^{\epsilon_p} \right]= c(n,p) \cdot \kappa_{p} \left[ \left(\frac{x}{n}\right)^{\epsilon_1}, \left(\frac{x}{n}\right)^{\epsilon_2}, \ldots,  \left(\frac{x}{n}\right)^{\epsilon_p} \right],
\end{equation}
where $c(n,p)$ is a constant satisfies 
\begin{equation*}
c(n,p)=
\begin{cases}
	n  &  \; \text{for odd} \; p, \\
	n^2  & \; \text{for even} \; p.
	\end{cases}
\end{equation*}
\end{rem}

In the rest of this section, we will provide some concrete examples for $\mu^\Gamma_x$. The following corollary was originally obtained by Nechita \cite{IN2018}.
\begin{cor}\label{cor:1} \cite{IN2018}.
	Let $x$ be a self-adjoint element in $(\mathcal{A},\varphi)$ with distribution $\mu_x$, and let $\mu_x^\Gamma$ be the distribution of $\tilde{x}^\Gamma$ Then we have 
	\begin{equation}\label{eq:Nechita}
	R_{\mu_x^\Gamma} (z) =\frac{1+n}{2}R_{\mu_x} \left(\frac{z}{n} \right)+\frac{1-n}{2}R_{\mu_x} \left(-\frac{z}{n} \right).
	\end{equation}
	Moreover, if $x$ is an even element, namely, all its odd moments vanish, then we have
	\[R_{\mu_x^\Gamma}(z)=n \cdot R_{\mu_x}\left( \frac{z}{n} \right).\]
\end{cor}

\begin{proof} By Theorem \ref{thm:1}, 
	\begin{align}
		\frac{1}{2} \left[ R_{\mu_x} \left(\frac{z}{n} \right)+R_{\mu_x} \left( -\frac{z}{n} \right) \right]
		& =\frac{1}{2}\sum_{p\geq0} \left[ \kappa_{p+1} [x,\ldots,x] \cdot \left( \frac{z}{n} \right)^p + \kappa_{p+1} [x,\ldots,x]\cdot \left(-\frac{z}{n} \right)^{p} \right] \notag\\
		& = \sum_{\text{even} \; p} \kappa_{p+1} [x,\ldots,x] \cdot \left( \frac{z}{n} \right)^{p}\notag\\
		& = \sum_{\text{even} \; p}n \cdot \kappa_{p+1}  \left[ \frac{x}{n},\ldots,\frac{x}{n} \right] \cdot z^{p}\notag\\
		& = \sum_{\text{even} \; p} \kappa_{p+1} \left[ \tilde{x}^\Gamma,\dots, \tilde{x}^\Gamma \right] \cdot z^{p},\notag
	\end{align}
	and similarly
	\begin{align*}
	\frac{n}{2} \left[ R_{\mu_x} \left( \frac{z}{n} \right)-R_{\mu_x} \left(-\frac{z}{n} \right) \right]
	=\sum_{\text{odd} \; p} \kappa_{p+1} \left[ \tilde{x}^\Gamma,\ldots, \tilde{x}^\Gamma \right] \cdot z^{p}.
	\end{align*}
Hence, 
	\begin{equation*}
		R_{\mu_x^\Gamma}(z) = \sum_{p\geq 0}  \kappa_{p+1} \left[ \tilde{x}^\Gamma,\ldots, \tilde{x}^\Gamma \right]  \cdot  z^{p} =\frac{1+n}{2}R_{\mu_x} \left( \frac{z}{n} \right)+\frac{1-n}{2}R_{\mu_x} \left( -\frac{z}{n} \right).
	\end{equation*}
Moreover, if $x$ is an even element, it is known that $R_{\mu_x}(z/n)+R_{\mu_x}(-z/n)=0$, which induces that $R_{\mu_x^\Gamma}(z)=n \cdot R_{\mu_x}\left( z/n \right)$.
\end{proof}

\noindent {\it The deterministic distribution}--
For any  $\alpha \in \R$, let $x=\alpha e$, the distribution of $x$ is given by $d\mu_x(t)=\delta_{\alpha}(t)dt$, and $R_{\mu_x}=\alpha.$
Then by Equation \ref{eq:Nechita} we have
\[R_{\mu_x^\Gamma}=\alpha= R_{\mu_x}. \]
It follows that $\mu_x = \mu_x^\Gamma.$ Let $y$ be another self-adjoint element in $\mathcal{A},$ which is free from $x$. Then we have
\begin{equation}\label{eq:distribution-Phi}
 \mu_{x+y}^{\Gamma}=\mu_{y}^{\Gamma}\boxplus\delta_{\alpha}.
\end{equation}

\vspace{2mm}

\noindent {\it Free Poisson and compound free Poisson distributions}--Let the parameter "rate" $\lambda\geq0$, and the parameter "jump size" $\alpha\in \R$. Let $\tilde{\mu}$ be a measure supported on the interval $[\alpha(1-\sqrt{\lambda})^{2},\alpha(1+\sqrt{\lambda})^{2}]$  with density
\begin{align*}
	d\tilde{\mu}(t)=\frac{1}{2\pi\alpha t}\sqrt{4\lambda\alpha^{2}-(t-\alpha(1+\lambda))^{2} }dt.
\end{align*}
The {\it free Poisson distribution} with rate $\lambda$ and jump size $\alpha$ \cite {NS2006} is given by 
\begin{equation}
\mu= 
\begin{cases}
	(1-\lambda)\delta_{0}+\lambda\tilde{\mu} &\text{if\ } 0\leq\lambda\leq1,\\
	\tilde{\mu}  &\text{if\ }\lambda>1,\\
\end{cases} \label{M-P-1}
\end{equation}
Let $\eta$ be a probability measure on $\mathbb{R}$ with compact support. Then the limit (in distribution) distribution $\nu$
    \[ \nu = \lim_{k \rightarrow \infty} \left[ \left(1-\frac{\lambda}{k} \right)\delta_{0}+\frac{\lambda}{k}\eta \right]^{\boxplus k}\]
    is called a {\it compound free Poisson distribution} with rate $\lambda$ and jump distribution $\eta$.

Now suppose that $x$ is a self-adjoint element in $(\mathcal{A},\varphi)$ with the free Poisson distribution $\mu_x$ (with rate $\lambda$ and jump size $\alpha$). Then its free cumulants and R-transform are given by $\kappa_{p}(\mu_x) =\lambda\alpha^{p}\, (p\geq1),$ and $R_{\mu_x}(z) =\sum_{p\geq0}\lambda\alpha^{p+1}z^{p}$ \cite{NS2006}. Hence
by Equation \ref{eq:Nechita} we have 
\begin{align*}
	R_{\mu_x^\Gamma}(z)
	&=\sum_{p\geq0}\left[\frac{1+n}{2}\lambda\alpha \left( \frac{\alpha}{n} \right)^{p}+\frac{1-n}{2}\lambda\alpha \left(-\frac{\alpha}{n} \right)^{p}\right]z^{p}\\
	&=\sum_{p\geq0}(\lambda n^{2})\left[\frac{1+n}{2n} \left( \frac{\alpha}{n} \right)^{p+1}+\frac{n-1}{2n} \left(-\frac{\alpha}{n} \right)^{p+1}\right]z^{p}\\
	&=\sum_{p\geq0}(\lambda n^{2})\cdot m_{p+1}(\eta)\cdot z^{p},
\end{align*}	
where the probability measure $\eta$ satisfies
\[d\eta(t)=\left[\frac{n+1}{2n}\delta_{\frac{\alpha}{n}}+\frac{n-1}{2n}\delta_{-\frac{\alpha}{n}}\right]dt.\] 
It follows that $R_{\mu_x^\Gamma}(z)=R_{\nu}(z)$, where $\nu$ is the following compound free Poisson distribution
\begin{align}\label{sp of compound poisson}
	 \nu = \lim_{k \rightarrow \infty} \left[ \left(1-\frac{\lambda n^{2}}{k} \right)\delta_{0}+\frac{\lambda n^{2}}{k}\eta \right]^{\boxplus k}.
\end{align}
Therefore $\mu_x^\Gamma = \nu.$


\section{Applications}

\noindent {\it $k$-positivity of linear maps and NPT problem}--Let $\Phi: \mathcal{M}_n(\mathbb{C}) \to \mathcal{M}_N(\mathbb{C})$ be a linear map. Its Choi matrix $C_\Phi  \in \M_n(\mathbb{C}) \otimes \M_N(\mathbb{C})$ is given by \cite{C75}
\begin{equation}\label{eq:Choi}
C_\Phi := \sum_{i,j=1}^n E_{ij} \otimes \Phi (E_{ij}). 
\end{equation}
On the other hand, for a given matrix $X \in \M_n(\mathbb{C}) \otimes \M_N(\mathbb{C}),$ we can associate a linear map $\Phi: \M_n(\mathbb{C}) \to \M_N(\mathbb{C})$ such that $X=C_\Phi$ via the following Choi-Jamiolkowksi isomorphism:
\begin{equation}\label{eq:map}
\Phi (A) : = {\rm Tr}_n \otimes \un \left[X \cdot A^\Gamma \otimes \un_N \right], \; \text{for all} \; A \in \M_n(\mathbb{C}),
\end{equation}
where $\Gamma$ is the transposition and ${\rm Tr}_n$ is the partial trace.

\begin{defn}\cite{C75}
\begin{enumerate}[{\rm (i)}]
\item A map $\Phi: \M_n(\mathbb{C}) \to \M_N(\mathbb{C})$ is called $k$-positive if the following dilation
$$\Phi \otimes \un_k: \M_n(\mathbb{C}) \otimes \M_k(\mathbb{C}) \to \M_N(\mathbb{C}) \otimes \M_k(\mathbb{C})$$
is positive. Moreover, it is called completely positive if it is $k$-positive for all $k,$ and it is co-completely positive if $\Gamma \circ \Phi$ is completely positive.
\item A matrix $X  \in \M_n(\mathbb{C}) \otimes \M_N(\mathbb{C})$ is called k-block positive if the matrix $(P\otimes \un_N) \cdot X \cdot (P\otimes \un_N)$ is positive for any rank $k$ orthogonal projection $P \in \M_n(\mathbb{C}).$
\end{enumerate} 
\end{defn}

It is well-known that $\Phi$ is completely positive if and only if $C_\Phi$ is positive. And for the $k$-positivity, we have the following result.
\begin{prop}\label{prop:block-positive}\cite{HLPQS15}.
Let $\Phi: \M_n(\mathbb{C}) \to \M_N(\mathbb{C})$ be a linear map, and its Choi matrix is $C_\Phi.$ The following are equivalent:
\begin{enumerate}[{\rm (a)}]
\item $\Phi$ is k-positive.
\item $C_\Phi$ is k-block positive.
\end{enumerate}
Note that the matrix algebra $\M_N(\mathbb{C})$ can be instead by the algebra of linear bounded operators on Hilbert space, which could be infinitely dimensional.
\end{prop}


In the entanglement theory, there is a strange phenomenon called "bound entanglement" \cite{HHH98}. 
The term "bound" means a lack of ability in quantum communication. This ability can be understood from the following distillability of the states \cite{BBPSSW96}: For a bipartite system, the state is distillable if one can (asymptotically) obtain maximally entangled states by local operations and classical communication (LOCC) from many copies of this state. Although the bound entanglement is not 
directly useful for quantum communication. However, it has been shown that the bound entangled states have positive effects on many other quantum tasks in an indirect way, especially for the production of a secure cryptographic key \cite{HHHO05, HPHH08, HHHO09, BCHW15}.     

Therefore, it is important to figure out the set of bound entanglements. It is well-known \cite{HHH97} that  for $2\otimes 2$ and $2 \otimes 3$ systems, all entangled states are distillable, thus there is no bound entangled state in such systems; however, starting from $3 \otimes 3$ and $2 \otimes 4$ systems, there already exist bound entangled states. The above results are based on the following famous positive partial transpose (PPT) criteria: Any PPT state is non-distillable, and thus any entangled PPT state is bound entangled \cite{HHH98}. Hence, it is natural to ask whether the converse statement is true. Alternatively, consider the following NPT problem:  
{\it Find bound entangled state with non-positive partial transposition (NPT).}  We refer to \cite{DSSTT2000, DCLB2000, BR03, PPHH2010} for some progress.  

There are two parallel lines of study for the NPT problem. One is to figure out an operational criteria for the distillability. It was shown \cite{HH99} that if NPT bound entangled states exist, then one can find such a state in a one parameter family of Werner states. Therefore using this idea, a promising approach was independently proposed by  two groups of people \cite{DSSTT2000} and \cite{DCLB2000} (see also \cite{BR03}). Recall a state $\rho$ is r-copy distillable if its r-copies tensor product $\rho^{\otimes r}$ can be locally projected to a $2\otimes 2$ NPT state, and $\rho$ is distillable if it is r-copy distillable for all $r.$ In their work, they focused on the r-copy distillability of Werner states, and showed that for any $r$, there is a range where the parameters are such that the Werner states are r-copy non-distillable. But unfortunately, the range becomes smaller as $r$ increases.    

In parallel, due to the Choi-Jamiolkowksi isomorphism, the NPT problem can be tackled by studying the $2$-positivity of tensor product of co-completely maps \cite{DSSTT2000, CTY18}. Namely, 
let $X \in \M_n(\mathbb{C}) \otimes \M_N(\mathbb{C})$ be a (block) matrix, and $\Phi: \M_n(\mathbb{C}) \to \M_N(\mathbb{C})$ be the linear map such that $X = C_\Phi.$ Then the co-complete positivity of $\Phi$ implies the positivity of 
$X^\Gamma,$ thus one can consider $X^\Gamma$ to be a (non-normalized) quantum state. Moreover, the $2$-positivity of $\Phi^{\otimes r}$ implies the $r$-copy non-distillability of $X^\Gamma.$
Thus, the existence of NPT bound entanglement is equivalent to the following \cite{MRW15, H06}: { \it Find a linear map $\Phi$ such that
(i) $\Phi$ is co-completely positive; (ii) $\Phi^{\otimes r}$ is 2-positive for all $r\geq 1$.}


\vspace{2mm}

\noindent {\it A new family of 1-copy non-distillable states with non-positive partial transposition}--In the rest of this section, the random (block) matrix $X_{nN} \in \M_n(\mathbb{C}) \otimes \M_N(\mathbb{C})$ is given as follows: 
\begin{equation}\label{eq:X}
X_{nN} = \un_{nN} + \alpha W_{nN},
\end{equation}
where $\alpha < 0$ and $W_{nN}$ is an $nN \times nN$ Wishart random matrix with parameter $\lambda \geq 0.$

\begin{prop}\label{prop:measure-X_N}
Almost surely as $N \to \infty,$
$X_{nN}$ strongly converges in distribution to a probability measure $\mu$ supporting on the interval $[1+\alpha(1+\sqrt{\lambda})^{2},1+\alpha(1-\sqrt{\lambda})^{2}] $ with density
\begin{align}
	d\mu(t)=\frac{1}{2\pi\alpha (t-1)}\sqrt{4\lambda\alpha^{2}-(t-1-\alpha(1+\lambda))^{2} }dt.
\end{align}
The probability measure $\mu$ is the translation of a free Poisson distribution with rate $\lambda$ and jump size $\alpha$.
\end{prop}

\begin{proof}
	By \cite[Corollary 2.2]{CM2012}, almost surely as $N \rightarrow \infty$, $\{W_{nN}, \un_{nN}\}$ strongly converges to $\{w, e\}$ , where $e$ is the unit in some $*$-probability space $(\mathcal{A}, \varphi)$, and $w \in \mathcal{A}$ is an element with the free poisson distribution with rate $\lambda$ and jump size $\alpha$, which is free with $e.$ Hence
	$X_{nN} = \un_{nN}+ \alpha W_{nN}$ strongly converges to $e+\alpha w.$ So  
	$$X_{nN} \xrightarrow{N \to \infty} \mu_{e+\alpha w},  \; \text{strongly convergence in distribution}.$$
 Thus by Equation \ref{eq:distribution-Phi}, $\mu_{e+\alpha w} = \delta_1\boxplus \mu_{\alpha w}$ and the density of $\mu_{e+\alpha w}$ is
	\begin{align}
		d\mu_{e+\alpha w}(t)=\frac{1}{2\pi\alpha (t-1)}\sqrt{4\lambda\alpha^{2}-(t-1-\alpha(1+\lambda))^{2}}dt.
	\end{align}
	with 
	${\rm supp}(\mu_{e+\alpha w})=[1+\alpha(1+\sqrt{\lambda})^{2},1+\alpha(1-\sqrt{\lambda})^{2}].$ We note that the above equation can also be obtained by functional calculus. 
\end{proof}

\begin{prop}\label{prop:measure-partial-transpose}
Let $X_{nN}^\Gamma$ be the partial transposition of $X_{nN}$, almost surely as $N \to \infty,$ $X_{nN}^\Gamma$ strongly converges in distribution to $\mu^\Gamma,$
where $\mu^\Gamma$ is given by the following compound free Poisson distribution
$$\mu^\Gamma= \lim_{k \rightarrow \infty} \left[ \left(1-\frac{\lambda n^{2}}{k} \right)\delta_{0}+\frac{\lambda n^{2}}{k}\eta \right]^{\boxplus k},$$
where the density of probability measure $\eta$ is given by
\[\eta(t)=\left[\frac{n+1}{2n}\delta_{\frac{\alpha}{n}}+\frac{n-1}{2n}\delta_{-\frac{\alpha}{n}}\right]dt.\] 
\end{prop}

\begin{proof}
It is obvious that $X_{nN}$ is Haar unitary invariant. Therefore, Theorem \ref{thm:PT} and Proposition \ref{prop:measure-X_N} combine to show that
$X_{nN}^\Gamma$ strongly converges in distribution to $\mu_{e+ \alpha w}^\Gamma.$ Hence, our proof follows from the computation in the previous section (see Equation \eqref{sp of compound poisson}).
\end{proof}

\begin{lem}\label{lemma4.4}
	Let $\{\mu_N\}_{N\geq 1}$ be a family of compactly supported measures such that
	$$\mu_N \xrightarrow{N \to \infty} \mu,  \; \text{convergence in distribution}.$$
	Moreover, if there exist a bounded interval $[a,b]\subseteq\R$, and a subsequence $\{N_{k}\}_{k\geq 1}$, such that 
	\[\rm{supp}(\mu_{N_{k}})\subseteq[a,b],\; \text{for all} \; k\geq1,\]
	then we have 
	\[\rm{supp}(\mu)\subseteq[a,b].\]
	\end{lem}
	
\begin{proof} Suppose that $a_{0}\in \text{supp}(\mu)$ and $a_{0}\in \R\backslash[a,b]$.  Without loss of generality, assume that $a_{0}<a$, and let $\epsilon= (a-a_0)/3>0$. Then there is an open neighborhood $\mathcal{O}$ of $a_0$ such that $\mathcal{O} \subseteq (a_{0}-\epsilon,a_{0}+\epsilon)$ and $\mu(\mathcal{O})>0$. 
		
		Let $f(t)$ be a non-negative function on $\R$ such that 
		\begin{equation*}
			f(t)=\begin{cases} 1, & \; \text{for} \; t\in(a_{0}-\epsilon,a_{0}+\epsilon); \\ 0, & \; \text{for} \; t\in(-\infty,a_{0}-2\epsilon)\cup(a_{0}+2\epsilon,+\infty), \end{cases}
		\end{equation*}
		thus by the convergence in distribution of $\{\mu_N\}_{N\geq 1}$ to $\mu,$ we have 
		\[0=\lim_{k\to \infty}\mu_{N_{k}}(f)=\mu(f)>0.\]
		A contradiction, which concludes our proof.
\end{proof}

\begin{prop}\label{prop:PPT}
Given $X_{nN}$ by Equation \eqref{eq:X}, assume that 
\begin{enumerate}[{\rm(i)}]
\item $\lambda>1$ and $\alpha<0$;
\item $1/n+2\sqrt{\lambda}+\lambda <-1/\alpha< (1+\sqrt{\lambda})^2,$
\end{enumerate}
then almost surely as $N \rightarrow \infty$, $X_{nN}^\Gamma$ is positive and $X_{nN}$ is non-positive.  
\end{prop}

\begin{proof}
Recall Proposition \ref{prop:measure-X_N}, the support of $\mu$ is given by 
			\[\text{supp}(\mu) = [\alpha(1+\sqrt{\lambda})^{2}+1,\alpha(1-\sqrt{\lambda})^{2}+1].\]
			Thus by the given conditions, we have 
			$$\alpha(1+\sqrt{\lambda})^{2}+1<0, \; \text{and} \; \alpha(1-\sqrt{\lambda})^{2}+1>0,$$ 
			which induces that $\inf \{\text{supp}(\mu)\} < 0.$
			By Proposition \ref{prop:measure-X_N}, $\mu$ is the strong limit in distribution of $\mu_N.$
			Thus by Lemma \ref{lemma4.4}, there is an $N_0\in \N,$ such that for all $N \geq N_0,$  
			\begin{equation*}
				\text{supp}(\mu_N)\not\subseteq[0,+\infty),
			\end{equation*}
			which concludes that $X_{nN}$ is non-positive.
			
On the other hand, recall that the distribution of $\mu^{\Gamma}$ is $\nu\boxplus\delta_{1}$, where $\nu$ is given as follow:
			\begin{align*}
				\nu =&\lim_{k \rightarrow \infty} \left[ (1-\frac{\lambda n^{2}}{k})\delta_{0}+\frac{\lambda n^{2}}{k}\tilde{\nu}\right]^{\boxplus k} \\
				=&\lim_{k \rightarrow \infty} \left[ (1-\frac{\lambda n^{2}}{k})\delta_{0}+\frac{\lambda n^{2}}{k}\bigg(\frac{n+1}{2n}\delta_{\frac{\alpha}{n}} +\frac{n-1}{2n}\delta_{-\frac{\alpha}{n}}\bigg) \right]^{\boxplus k}\\
				:= & \lim_{k \rightarrow \infty} \nu_k^{\boxplus k}, 
			\end{align*}
			where $\nu_k = (1-\frac{\lambda n^{2}}{k})\delta_{0}+\frac{\lambda n^{2}}{k}\tilde{\nu}.$ It is easy to see that the mean $m$ and variance $\sigma^{2}$ of $\nu_k$ is given by  
			\begin{align*}
				m=\frac{\lambda\alpha}{k}, \; \text{and} \; \sigma^{2}=\frac{\lambda\alpha^{2}}{k}- \left( \frac{\lambda\alpha}{k} \right)^{2}.
			\end{align*}
			By Proposition \ref{prop:est-free-convolution-power} (\cite[Lemma 2.3]{BM2015}), we have the following estimation of supp$(\nu_k^{\boxplus k})$
			\begin{align*}
				\inf \left\{ \text{supp} \left(\nu_k^{\boxplus k} \right) \right\} &\geq \frac{\alpha}{n}+ (k-1) m -2 \sigma \sqrt{k-1}\\
				& =\frac{\alpha}{n}+\frac{\lambda\alpha(k-1)}{k}-2\sqrt{\frac{\lambda\alpha^{2}(k-\lambda)(k-1)}{k^{2}}}\\
				&>\frac{\alpha}{n}+\lambda\alpha-2\sqrt{\lambda\alpha^{2}}.
			\end{align*}
			
			Moreover, by the given condition we have $\alpha/n+\lambda\alpha-2\sqrt{\lambda\alpha^{2}}+1>0$. Hence by Lemma \ref{lemma4.4}, we have
			\begin{align*}
				\inf\left\{ \text{supp}(\mu^{\Gamma}) \right\}& = 1+\inf\left\{ \text{supp}(\nu) \right\}\\
				&\geq 1+\inf_{k} \left\{ \inf \left\{ \text{supp} \left( \nu_k^{\boxplus k} \right) \right\} \right\}\\
				&\geq 1+\frac{\alpha}{n}+\lambda\alpha-2\sqrt{\lambda\alpha^{2}}>0,
			\end{align*}
			which induces that $\inf\{\text{supp}(\mu^{\Gamma}) \}> 0.$ Again by Proposition \ref{prop:measure-partial-transpose}, for sufficiently large $N$ we have $X_{nN}^{\Gamma}$ is positive. 
\end{proof}

\begin{prop}\label{prop:k-positive}
Let $X_{nN}$ be given in Proposition \ref{prop:PPT}, for integers $k\geq 1,$ almost surely one has: for sufficiently large $n$ and $N$, $X_{nN}$ is $k$-block positive. 
\end{prop}

\begin{proof}
Firstly we claim the following: If ${\rm supp} (\mu^{\boxplus n/k}) \subset (0, \infty)$, then almost surely as $N \rightarrow \infty,$ $X_{nN}$ is $k$-block positive. Note that this claim is a result in \cite{CHN2016}; however, we provide a proof for completeness. 

For a fixed rank $k$ projection $Q \in \M_n (\mathbb{C})$, by Proposition \ref{prop:strong} (\cite[Theorem 4.1, Lemma 3.1]{CHN2016}), almost surely
we have
\begin{equation*}
\lim_{N \rightarrow \infty} \lambda_{\min}\left[ (Q \otimes \un_{N}) \cdot X_{nN} \cdot (Q \otimes \un_{N}) \right] = \frac{k}{n} \inf \left( {\rm supp} (\mu^{\boxplus n/k}) \right) \geq \frac{k \delta}{n}.  
\end{equation*}
For all rank $k$ projection $P \in \M_n (\mathbb{C}),$ there exists a rank $k$ projection $Q\in \M_n (\mathbb{C})$ such that $\|P-Q\|< \epsilon.$ Moreover, almost surely we have
$$\|(P \otimes \un_{N}) \cdot X_{nN} \cdot (P\otimes \un_{N}) - (Q \otimes \un_{N}) \cdot X_{nN} \cdot (Q \otimes \un_{N})\| \leq \|P-Q\| \cdot \|X_{nN}\| \leq M\epsilon,$$
where $M> \max \left( {\rm supp} (\mu) \right)= \|X_{nN}\|$ is a constant. Therefore by using the triangle inequality, we can conclude that, almost surely, 
$$\lim_{N \rightarrow \infty} \min_{P} \lambda_{\min} \left[ (P \otimes \un_{N}) \cdot X_{nN} \cdot (P \otimes \un_{N}) \right] \geq \frac{k\delta}{n}-M\epsilon.$$
Hence by letting $\epsilon< (k\delta)/(n M)$ and using Proposition \ref{prop:block-positive} we can conclude our claim.

Recall that the support of $\mu$ is given as follows:
	\[{\rm supp}(\mu)=[1+\alpha(1+\sqrt{\lambda})^{2},1+\alpha(1-\sqrt{\lambda})^{2}].\]
Denote $m$ and $\sigma^{2}$ by the mean and variance of $\mu$ respectively, then we have 
	\begin{align*}
		m =\lambda\alpha+1, \; \text{and} \; \sigma^{2}= \lambda\alpha^2.
	\end{align*}
Again by Proposition \ref{prop:est-free-convolution-power} (\cite[Lemma 2.3]{BM2015}), we have the following estimation of ${\rm supp} (\mu^{\boxplus n/k})$
	\begin{equation}\label{eq:positive-condition}
	\begin{split}
		\inf\left\{ {\rm supp}(\mu^{\boxplus n/k})\right\} 
		& \geq\left(\alpha(1+\sqrt{\lambda})^{2}+1\right) + \left(\frac{n}{k}-1 \right)m-2\sigma \sqrt{\frac{n}{k}-1}\\
		& := f(n).
		\end{split}
	\end{equation}

Note that the conditions for the parameters $\alpha$ and $\lambda$ induces that 
$$m >0, \; \text{and} \; \alpha(1+\sqrt{\lambda})^{2}+1 <0.$$
Hence for any integer $k \geq 1$, we can always find a sufficient large $n_0,$ such that for any $n \geq n_0,$ $f(n)>0.$ The rest of the proof follows the claim.
\end{proof}

In summary, we explicitly construct a family of random linear maps $\{\Phi_N\}_N,$ such that Choi matrix of $\Phi_N$ is $X_{nN}$ (see Equations \eqref{eq:Choi} and \eqref{eq:map}).
For any integer $k\geq 1$, suppose that the parameters $\alpha$ and $\lambda$ fulfill the conditions in Proposition \ref{prop:PPT}, then almost surely one has: for sufficiently large $N$ and $n$, 
\begin{enumerate}[(a)]
\item $\Phi_N$ is co-completely positive, while $\Phi_N$ is not completely positive;
\item $\Phi_N$ is $k$-positive.
\end{enumerate}
By letting $k=2,$ a direct corollary of our result is that we have explicitly obtained a new family ($X_{nN}^\Gamma$) of $1$-copy non-distillable states with non-positive partial transposition. Hence, 
a possible way to solve the NPT problem is to consider the limit distribution of $X_{nN}^{\otimes r}$ (or $(X_{nN}^\Gamma)^{\otimes r}$) for all $r \geq 1.$

\vspace{2mm}

{\it Acknowledgments}--We would like to thank Benoit Collins for his very stimulating and fruitful discussions.
We are partially supported by NSFC No. 12031004. Liang is partially supported by CSC No. 202006120259.



\bibliography{Limit-distribution-of-partial-transposition-of-block-random-matrices}

\end{document}